\documentclass[hidelinks,12pt]{article}
\usepackage[export]{adjustbox} 
\usepackage{scicite}
\usepackage{color}
\usepackage{graphicx}	
\usepackage{amssymb}
\usepackage{enumitem}
\usepackage{bm}
\usepackage{caption}
\usepackage{subcaption}
\usepackage{mathtools}
\usepackage{titlesec}
\usepackage{bbm}

\usepackage{colortbl}
\definecolor{Gray}{gray}{0.9}
\usepackage{hyperref}
\makeatletter
\renewcommand*{\@biblabel}[1]{\hfill#1.}
\makeatother
\newcounter{myequation}
\makeatletter
\@addtoreset{equation}{myequation}
\makeatother
\makeatletter
\renewcommand*{\@textcolor}[3]{%
\protect\leavevmode
\begingroup
\color#1{#2}#3%
\endgroup
}
\makeatother

\usepackage{times}
\emergencystretch=\maxdimen
\usepackage[labelsep=period]{caption}
\usepackage[labelfont=bf]{caption}
\usepackage{amsmath,accents}
\usepackage[para,online,flushleft]{threeparttable}
\usepackage{multirow}
\usepackage{xltabular}
\usepackage{threeparttablex}
\usepackage{booktabs, caption}
\usepackage{makecell} 

\newenvironment{sciabstract}{%
\begin{quote} \bf}
{\end{quote}}

\renewcommand{\figurename}{Fig.}
\usepackage[table]{xcolor}

\title{\textbf{Learning Ordering in Crystalline Materials with Symmetry-Aware Graph Neural Networks}}

\author{
\hspace{0mm}Jiayu Peng,$^{1,\dag}$ James Damewood,$^{1,2,\dag}$ Jessica Karaguesian,$^{1,2,\dag}$ Jaclyn R.\\Lunger,$^{1}$ Rafael Gómez-Bombarelli$^{1,*}$\\
\\
\hspace{0mm}\normalsize{$^{1}$Department of Materials Science and Engineering,}\\
\hspace{0mm}\normalsize{$^{2}$Center for Computational Science and Engineering,}\\
\hspace{0mm}\normalsize{Massachusetts Institute of Technology, Cambridge, MA 02139, USA}\\
\hspace{0mm}{$^{\dag}$\small Equal contribution; $^{*}$\small rafagb@mit.edu (R.G.-B.)}\\
}

\date{}

\begin{document}


\maketitle 

\baselineskip18pt

\phantomsection
\addcontentsline{toc}{section}{Abstract}
\begin{sciabstract}

Graph convolutional neural networks (GCNNs) have become a machine learning workhorse for screening the chemical space of crystalline materials in fields such as catalysis and energy storage, by predicting properties from structures. Multicomponent materials, however, present a unique challenge since they can exhibit chemical (dis)order, where a given lattice structure can encompass a variety of elemental arrangements ranging from highly ordered structures to fully disordered solid solutions. Critically, properties like stability, strength, and catalytic performance depend not only on structures but also on orderings. To enable rigorous materials design, it is thus critical to ensure GCNNs are capable of distinguishing among atomic orderings. However, the ordering-aware capability of GCNNs has been poorly understood. Here, we benchmark various neural network architectures for capturing the ordering-dependent energetics of multicomponent materials in a custom-made dataset generated with high-throughput atomistic simulations. Conventional symmetry-invariant GCNNs were found unable to discern the structural difference between the diverse symmetrically inequivalent atomic orderings of the same material, while symmetry-equivariant model architectures could inherently preserve and differentiate the distinct crystallographic symmetries of various orderings.

\end{sciabstract}

\clearpage

\phantomsection
\addcontentsline{toc}{section}{Introduction}
\section*{Introduction}

Graph convolutional neural networks (GCNNs) have been widely leveraged as the state-of-the-art machine learning (ML) models for learning about solid materials at the atomic scale \cite{Reiser:2022,Damewood:2023,Schutt:2018,Xie:2018,Chen:2019,Schutt:2021,Geiger:2022,Batzner:2022,Batatia:2022} by mapping atomic structures to property labels in diverse applications, ranging from designing superalloys \cite{Morgan:2020} to discovering uncharted materials for catalyzing sustainability and decarbonization \cite{Peng:2022}. These models learn materials properties through message passing and graph convolution among atoms and their coordination environments in crystal structures \cite{Reiser:2022,Damewood:2023,Schutt:2018,Xie:2018,Chen:2019,Schutt:2021,Geiger:2022,Batzner:2022,Batatia:2022}. For example, GCNNs have been used to alleviate the formidable cost of high-throughput first-principles atomistic simulations for screening multicomponent oxide catalysts \cite{Lunger:2024} and infer defect formation energetics \cite{Witman:2023}. More recently, GCNNs have been shown to have great promise for serving as foundation models for computational materials discovery \cite{Merchant:2023,Batatia:2024}, e.g., in the form of universal interatomic potentials, which have low computational cost but retain ab initio accuracy across the periodic table for inexpensive, accurate simulations on unprecedented length and time scales \cite{Chen:2022,Deng:2023}. Moreover, GCNNs have also been demonstrated as crucial building blocks for creating generative models to efficiently discover new crystalline materials out of an unlimited design space \cite{Zeni:2024}.

Given the wide use of GCNNs for materials design, it is critical to understand their limitations and how they may be addressed through architectural choices. For example, GCNNs are well known to be data-intensive, requiring at least thousands of data points to train a prediction model \cite{Reiser:2022,Peng:2022,Lunger:2024,Witman:2023} or up to millions to build a universal interatomic potential \cite{Merchant:2023,Batatia:2024,Chen:2022,Deng:2023}. Moreover, the underlying mathematical frameworks of GCNNs have been shown to result in the limitations of oversmoothing \cite{Keriven:2022} and oversquashing \cite{Gong:2023}. Each graph convolutional layer in GCNNs behaves like a graph-smoothing operator that effectively homogenizes the information encoded on different atoms, which can constrain the performance of models with many convolutional layers \cite{Keriven:2022}. GCNNs also fall short in learning the periodicity of crystal structures, as the graph convolutional layers focus on encoding local coordination environments, instead of capturing long-range interactions or extensive materials properties \cite{Gong:2023}.

GCNNs have been shown to capture the dependence of materials properties on chemical compositions \cite{Reiser:2022,Lunger:2024,Witman:2023,Merchant:2023,Batatia:2024,Chen:2022,Deng:2023,Zeni:2024} (Fig. \ref{fig:Main_knowledge_gap}a), but there is still a lack of understanding of their ability to capture and differentiate complex atomic orderings in crystalline materials. In addition to compositional tunability, crystalline materials can have diverse forms of chemical disorder (i.e., the distribution of different chemical elements within a crystal structure) and structural disorder (i.e., the deviation of a crystal structure from a perfectly crystalline lattice). In other words, materials with the highest degree of chemical or structural disorder are ideal solid solutions or amorphous solids, respectively. Chemical (dis)order, that is, control of the atomic ordering in crystalline materials, has been widely leveraged to optimize properties, e.g., catalytic activity \cite{Liu:2023}, ion conductivity \cite{Zeng:2022}, and magnetic \cite{Kumar:2021}, optical \cite{Nechache:2015}, and electrochemical performance \cite{Lun:2021}. Chemical ordering can be modeled by conventional computational approaches, in particular using special quasirandom structures \cite{Zunger:1990} to mimic long-range disorder in midsize supercells together with cluster expansions of the total energy \cite{Yang:2022} to predict energies of arbitrary orderings. However, such traditional approaches are too costly to explore the union of possible atomic orderings and potential chemical compositions across large chemical space \cite{Peng:2024}. In contrast, GCNNs could be a low-cost surrogate to score the energies and properties of various possible orderings. Unfortunately, a systematic benchmark is still missing on the capability of such GCNN-based representation learning to distinguish diverse possible orderings from one another and capture the ordering dependence of materials properties (Fig. \ref{fig:Main_knowledge_gap}a). This knowledge gap is evidenced in a recent controversy over the neglect of chemical (dis)order in GCNN-powered computational materials screening \cite{Merchant:2023,Cheetham:2024}. This limitation may be tackled by examining an extended set of possible orderings with GCNNs to estimate the thermodynamic tendency to form chemically disordered solid solutions \cite{Peng:2024}. To this end, it is vital to evaluate if GCNNs can learn the dependence of materials properties on compositions and orderings simultaneously.

We hypothesize that distinguishing the atomic orderings of crystalline materials by GCNNs requires symmetry-aware models that can encode and contrast the distinct symmetries of crystal structures with the same composition but different orderings. Symmetry-invariant GCNNs, such as SchNet \cite{Schutt:2018}, CGCNN \cite{Xie:2018}, and MEGNet \cite{Chen:2019}, characterize chemical coordination environments in crystal structures as scalars (Fig. \ref{fig:Main_compositional_dependence}a). Such models are invariant to the Euclidean group E(3), i.e., all the symmetry operations of translations, rotations, and reflections in a three-dimensional (3D) space. In contrast to these invariant models \cite{Schutt:2018,Xie:2018,Chen:2019}, a new class of GCNNs, e.g., PaiNN \cite{Schutt:2021}, e3nn \cite{Geiger:2022}, NequIP \cite{Batzner:2022}, and MACE \cite{Batatia:2022}, have been developed recently to encode local coordination environments as vectors or higher-order tensors, such that their representations are equivariant to translations, rotations, and reflections (Fig. \ref{fig:Main_compositional_dependence}a). These equivariant GCNNs \cite{Schutt:2021,Geiger:2022,Batzner:2022,Batatia:2022} inherently preserve and capture the crystallographic symmetries of input crystal structures \cite{Smidt:2021}, making them more data-efficient than their symmetry-invariant counterparts \cite{Chen:2021}. Notably, equivariant GCNNs have been recently utilized to quantify symmetry breaking in physical systems \cite{Sheriff:2024,Wang:2024} and to predict symmetry-constrained properties \cite{Wen:2024}. However, it remains unclear whether GCNNs can effectively differentiate symmetrically inequivalent atomic orderings in crystalline materials and accurately learn both the compositional and ordering dependence of their properties.

In this study, combining representation learning and high-throughput density functional theory (DFT) atomistic simulations, we systematically assessed symmetry-invariant and symmetry-equivariant GCNNs and benchmarked their capability to encode and differentiate atomic orderings in solid materials. We built a DFT dataset composed of the structures and energies of over $10,000$ multicomponent perovskite oxides with various cation orderings and further compared CGCNN \cite{Xie:2018} and e3nn \cite{Geiger:2022} as representative invariant and equivariant models, respectively. We found that while both types of GCNNs can capture the composition-dependent thermodynamic stability of multicomponent perovskite oxides, only symmetry-equivariant models can characterize the ordering dependence of ab initio thermodynamic stability. Moreover, through latent embedding analysis, we showed that symmetry-invariant GCNNs fall short in differentiating various symmetrically inequivalent orderings through message passing and graph convolution. On the contrary, equivariant GCNNs can better distinguish such orderings from one another by encoding crystallographic symmetries into their model architectures. We observed that the difference between invariant and equivariant GCNNs in capturing orderings can be demonstrated across a broad range of perovskite compositions and model architectures. Our findings emphasize the value of symmetry-aware ML models for effectively understanding and accurately predicting atomic orderings in crystalline materials.

\phantomsection
\addcontentsline{toc}{section}{Results}
\section*{Results}

\phantomsection
\addcontentsline{toc}{subsection}{Building a high-throughput dataset of cation orderings in perovskite oxides}
\subsection*{Building a high-throughput dataset of cation orderings in perovskite oxides}

\begin{figure}
\phantomsection
\begin{center}
\includegraphics[max size={\textwidth}{\textheight}]{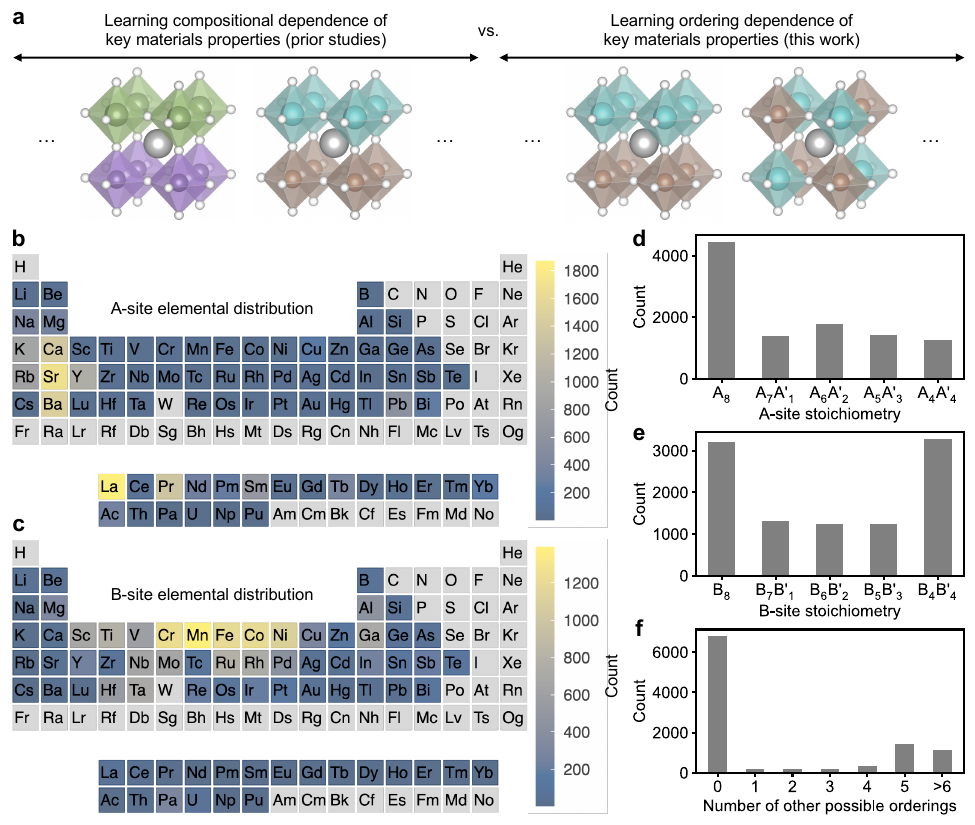}
\caption{
\textbf{Existing knowledge gap and high-throughput DFT dataset.}
While GCNNs have been widely shown to learn the dependence of materials properties on chemical compositions, it is unknown to what degree they can capture the dependence on atomic orderings. Color scheme: A, grey; B, cyan or green; B', brown or purple; O, white. \textbf{b},\textbf{c}, Occurrences of A-site (\textbf{b}) and B-site elements (\textbf{c}) in our high-throughput DFT dataset of multicomponent perovskite oxides (Table \ref{table:dataset_comparison}), which encompasses a variety of elements, including (but not limited to) commonly observed A-site (e.g., alkaline earth and rare earth metals) and B-site cations (e.g., transition metals), suggesting that this dataset covers compositions across a broad space of multicomponent perovskite oxides. \textbf{d}--\textbf{f}, Histograms for the occurrences of A-site (\textbf{d}) and B-site stoichiometries (\textbf{e}) and the numbers of other symmetrically inequivalent cation orderings (\textbf{f}) examined computationally for all oxide structures in our dataset. As the occurrences of A\textsubscript{\textit{x}}A'\textsubscript{8--\textit{x}} and B\textsubscript{\textit{y}}B'\textsubscript{8--\textit{y}} are equal to those of A\textsubscript{8--\textit{x}}A'\textsubscript{\textit{x}} and B\textsubscript{8--\textit{y}}B'\textsubscript{\textit{y}}, respectively, we only showed the stoichiometries of $8 \geq x, y \geq 4$ (\textbf{d},\textbf{e}) for clarity.
}
\label{fig:Main_knowledge_gap}
\end{center}
\end{figure}

We selected multicomponent perovskite oxides as the materials space of interest and built a dataset of over $10,000$ pairs of relaxed structures and matching energies through high-throughput DFT atomistic simulations. We focused on perovskite oxides due to their compositional and structural flexibility. This dataset comprises perovskites with a formula of A\textsubscript{\textit{x}}A'\textsubscript{1--\textit{x}}B\textsubscript{\textit{y}}B'\textsubscript{1--\textit{y}}O\textsubscript{3}. DFT calculations were performed on supercells containing 40 atoms, such that $x$ and $y$ have a range between $0$ and $1$ in $0.125$ intervals. Structurally, perovskite oxides consist of a network of corner-shared BO\textsubscript{6} octahedra surrounded by A-site cations. These structures can host a large set of A-site and B-site cations from across the periodic table, giving rise to various stoichiometries and cation orderings within our high-throughput DFT dataset (Fig. \ref{fig:Main_knowledge_gap}b--f). When compared with previous high-throughput DFT studies on perovskite oxides \cite{Castelli:2012,Emery:2017,Korbel:2016,Schmidt:2017,Jacobs:2018,Talapatra:2021,Ma:2021,Wang:2022,Bare:2023,Wang2:2024}, our dataset is much more diverse both compositionally and structurally, especially in cation orderings (Table \ref{table:dataset_comparison}), offering a balance between the breadth and depth of DFT data for training GCNNs and assessing whether these models can learn their dependence on compositions and orderings concurrently.

As the GCNN models to be evaluated act on 3D crystal structures, the choice of input structures is critical. The most informative structures to predict DFT-level energies are the corresponding DFT-relaxed structures, but those would not be freely accessible in a high-throughput screening setting\cite{Peng:2024}. In addition to these information-rich structures that are ultimately unavailable at inference time, we evaluated lower-grade crystal geometries as inputs, effectively convolving the energy prediction task with an implicit geometry refinement one, similar to the initial structure to relaxed energy (IS2RE) task in the Open Catalyst Project \cite{Chanussot:2021} and Matbench Discovery \cite{Riebesell:2024}. Specifically, we evaluate the use of idealized unrelaxed, high-symmetry cubic geometries that can be generated in milliseconds and, in principle, contain all the information for learning energies, since there is only one global energy minima for each ordering. Realistically, perovskite structures likely have much lower symmetries than being idealized cubic owing to spontaneous symmetry breaking (Fig. \ref{fig:Main_compositional_dependence}a), such as octahedral tilting and Jahn--Teller distortions \cite{King:2010}. As a result, a trade-off is expected between the structural fidelity and computational expense of input data when different types of perovskite structures are utilized \cite{Law:2023,Li:2024}. A middle ground can be potentially achieved by structures relaxation using empirical methods \cite{Lufaso:2006}, ML potentials \cite{Merchant:2023,Batatia:2024,Chen:2022,Deng:2023} or generative models \cite{Zeni:2024} which are accessible at a cost of roughly seconds to minutes per structure in graphics processing unit (GPU) time.

\phantomsection
\addcontentsline{toc}{subsection}{Predicting compositional dependence of ab initio thermodynamic stability}
\subsection*{Predicting compositional dependence of ab initio thermodynamic stability}

\begin{figure}
\phantomsection
\begin{center}
\includegraphics[max size={\textwidth}{\textheight}]{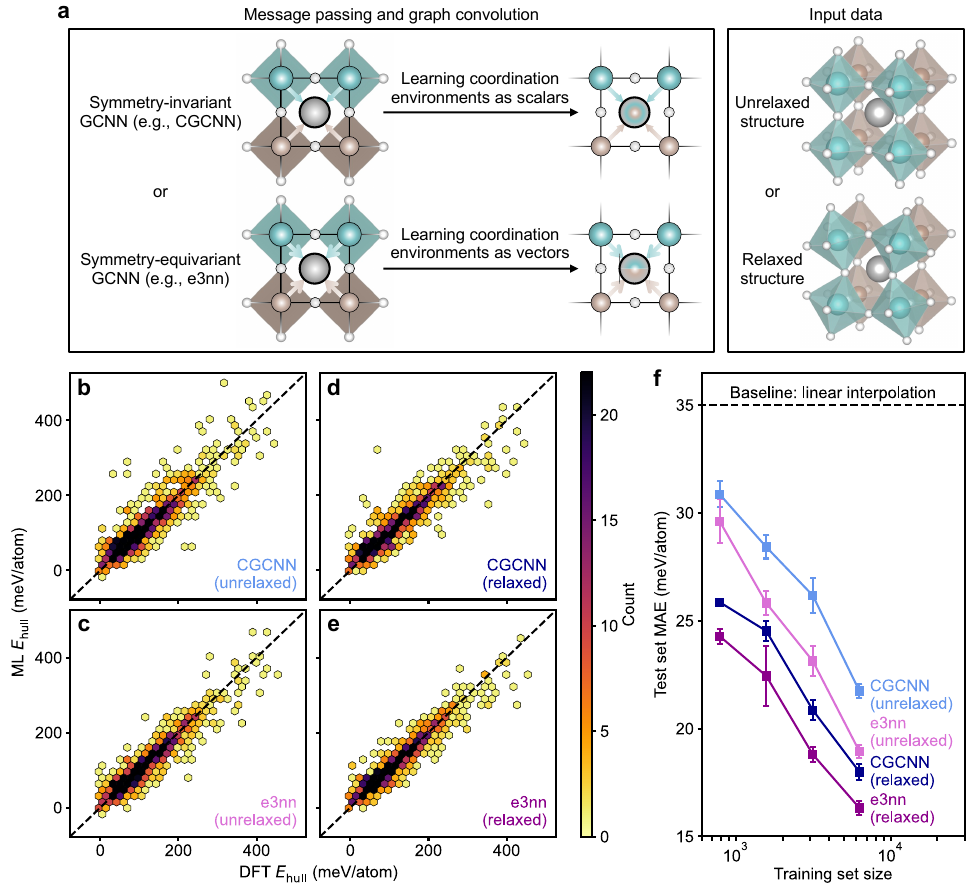}
\caption{
\textbf{Representation learning with symmetry-invariant and symmetry-equivariant GCNNs across compositional space.}
\textbf{a}, Message passing and graph convolution in invariant and equivariant GCNNs, where the information of coordination environments is encoded as scalars and vectors (or higher-order tensors), respectively. Regarding input perovskite oxide data, unrelaxed, idealized structures result in minimum computational cost yet low data fidelity, while DFT-relaxed, distorted perovskite structures possess higher data fidelity but also greater computational expense. Color scheme: A, grey; B, cyan; B', brown; O, white. \textbf{b}--\textbf{e}, GCNN-predicted vs. DFT-computed $E_\mathrm{hull}$ for a test set of $1,261$ perovskite oxides, where CGCNN and e3nn were trained on either unrelaxed (\textbf{b},\textbf{c}) or DFT-relaxed structures (\textbf{d},\textbf{e}). \textbf{f}, Test set MAE as a function of training set size. The baseline implies the MAE from simply predicting $E_\mathrm{hull}$ as a linear interpolation of the DFT-computed $E_\mathrm{hull}$ of ternary perovskites. All data points and error bars denote the mean and standard deviation of the three best-performing models, respectively.
}
\label{fig:Main_compositional_dependence}
\end{center}
\end{figure}

We trained invariant and equivariant GCNNs to predict the energy above the convex hull ($E_\mathrm{hull}$) of our multicomponent perovskite oxide dataset. $E_\mathrm{hull}$ is defined as the energetic driving force for the decomposition of a material into competing phases \cite{Bartel:2020}, which characterizes thermodynamic stability with respect to decomposition and has been widely leveraged as a metric to estimate the success of ML-driven computational materials discovery \cite{Lunger:2024,Witman:2023,Merchant:2023,Batatia:2024,Chen:2022,Zeni:2024,Ma:2021,Law:2023}. For multicomponent perovskite oxides, apart from decomposition to the competing phases, it is also essential to predict their thermodynamic stability with respect to alternative orderings at the same composition. Our recent work \cite{Peng:2024} has shown that the relative DFT-computed energies of various symmetrically distinct cation arrangements in multicomponent perovskites are well correlated with their experimental cation orderings, highlighting that $E_\mathrm{hull}$ captures the thermodynamic tendency to give rise to alternative orderings in multicomponent perovskite oxides.

Both symmetry-invariant and symmetry-equivariant GCNNs were found to capture accurately the composition-dependent thermodynamic stability of multicomponent perovskite oxides (Fig. \ref{fig:Main_compositional_dependence}). When trained and evaluated on unrelaxed perovskite structures (Fig. \ref{fig:Main_compositional_dependence}a), symmetry-invariant CGCNN shows a mean absolute error (MAE) of $21.7\ \mathrm{meV/atom}$ (Fig. \ref{fig:Main_compositional_dependence}b), whereas symmetry-equivariant e3nn has a moderately lower MAE of $19.0\ \mathrm{meV/atom}$ (Fig. \ref{fig:Main_compositional_dependence}c). Switching inputs to relaxed structures further decreases the MAEs to $18.0\ \mathrm{meV/atom}$ (Fig. \ref{fig:Main_compositional_dependence}d) and $16.3\ \mathrm{meV/atom}$ (Fig. \ref{fig:Main_compositional_dependence}e) for the invariant CGCNN and equivariant e3nn, respectively. These trends are consistent with recent studies suggesting that using equivariant model architectures \cite{Schutt:2021,Batzner:2022,Batatia:2022} or DFT-relaxed input structures \cite{Law:2023,Li:2024} boosts the performance of GCNNs. Prior work on composition-only ML models that do not use structural information has reported an MAE of more than $120\ \mathrm{meV/atom}$ when trained on the $E_\mathrm{hull}$ of $20,000$ five-atom perovskite unit cells \cite{Schmidt:2017}, which is considerably larger than those of GCNNs with a much smaller training set (Fig. \ref{fig:Main_compositional_dependence}f), in agreement with recent benchmarks showing that GCNNs outperform composition-only models for learning composition-dependent thermodynamic stability \cite{Bartel:2020}. The MAEs of CGCNN and e3nn are also significantly smaller than that of a baseline prediction method ($35\ \mathrm{meV/atom}$; Fig. \ref{fig:Main_compositional_dependence}f) that assumes the $E_\mathrm{hull}$ of multicomponent perovskite oxides (such as A\textsubscript{\textit{x}}A'\textsubscript{1--\textit{x}}B\textsubscript{\textit{y}}B'\textsubscript{1--\textit{y}}O\textsubscript{3}) simply as a linear interpolation of the DFT-computed $E_\mathrm{hull}$ of corresponding ternary perovskites (i.e., ABO\textsubscript{3}, A'BO\textsubscript{3}, AB'O\textsubscript{3}, and A'B'O\textsubscript{3}).

\phantomsection
\addcontentsline{toc}{subsection}{Distinguishing orderings through symmetry-aware graph neural networks}
\subsection*{Distinguishing orderings through symmetry-aware graph neural networks}

\begin{figure}
\phantomsection
\begin{center}
\includegraphics[max size={\textwidth}{\textheight}]{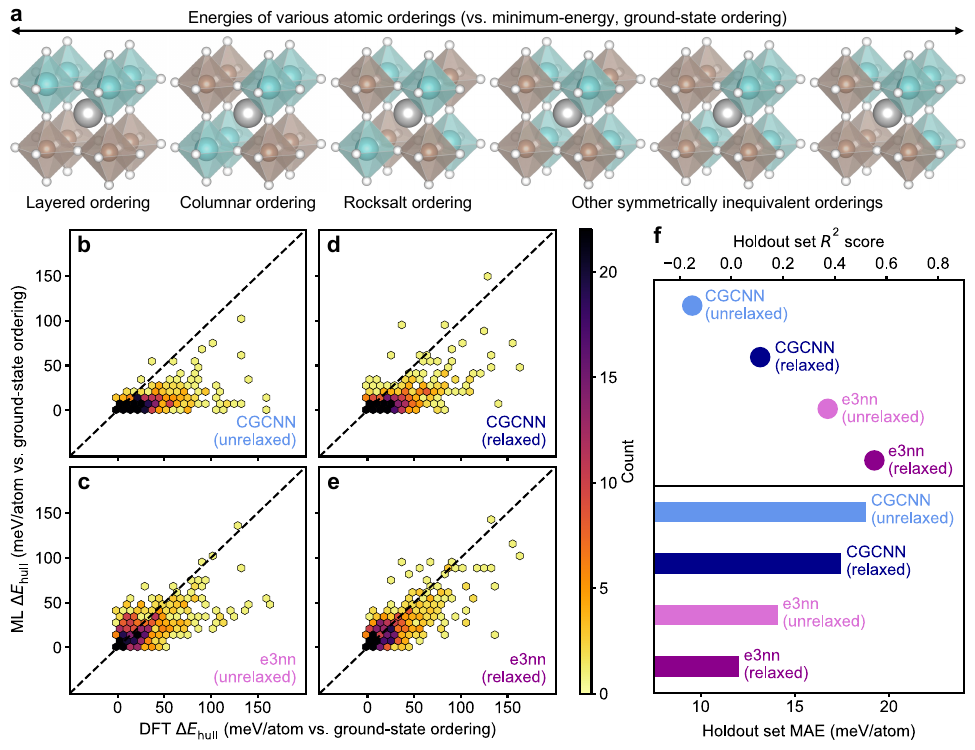}
\caption{
\textbf{Representation learning with symmetry-invariant and symmetry-equivariant GCNNs across ordering space.}
\textbf{a}, All possible symmetrically unique cation arrangements in an idealized cubic supercell of AB\textsubscript{0.5}B'\textsubscript{0.5}O\textsubscript{3} perovskites with 40 atoms, when distributing four B and four B' cations among the eight B-site locations in the supercell. Color scheme: A, grey; B, cyan; B', brown; O, white. The relative DFT-computed energies of these symmetrically unique cation arrangements can be well correlated with the experimental orderings of AB\textsubscript{0.5}B'\textsubscript{0.5}O\textsubscript{3} across a broad oxide space \cite{Peng:2024}. \textbf{b}--\textbf{f}, GCNN-predicted vs. DFT-computed $E_\mathrm{hull}$ of all symmetrically unique cation arrangements relative to those of GCNN-predicted or DFT-computed ground-state arrangements, respectively, for a holdout set of $100$ AB\textsubscript{0.5}B'\textsubscript{0.5}O\textsubscript{3} perovskite compositions, where each composition has six symmetrically unique arrangements (\textbf{a}), and CGCNN and e3nn were trained on either unrelaxed (\textbf{b},\textbf{c}) or DFT-relaxed structures (\textbf{d},\textbf{e}), with holdout set $R^2$ score and MAE shown for all models and inputs (\textbf{f}). All data points denote the mean of the three best-performing models.
}
\label{fig:Main_ordering_dependence}
\end{center}
\end{figure}

We then focused on predicting the ordering-dependent $E_\mathrm{hull}$ of symmetrically inequivalent cation arrangements (Fig. \ref{fig:Main_ordering_dependence}a) in AB\textsubscript{0.5}B'\textsubscript{0.5}O\textsubscript{3} perovskite oxides. We particularly assessed these compositions because they have dominated the existing literature with experimentally quantified cation orderings \cite{King:2010}. Regarding the six symmetrically distinct cation arrangements possible in a supercell of 40 atoms (Fig. \ref{fig:Main_ordering_dependence}a), our recent work has shown that when all orderings have similar energies, the perovskite oxides tend to be experimentally cation-disordered, whereas when one particular arrangement is much lower in energy than the other (by, e.g., $<20\ \mathrm{meV/atom}$), the perovskite oxides are most likely experimentally cation-ordered \cite{Peng:2024}. Therefore, the relative energy difference between $E_\mathrm{hull}$ of various cation arrangements vs. the lowest-energy one can serve as an accurate descriptor of cation orderings in multicomponent perovskite oxides\cite{Peng:2024}.

We found that symmetry-invariant GCNNs poorly characterize the ordering-dependent thermodynamic stability of multicomponent perovskite oxides, but equivariant GCNNs perform much better in learning the energy difference between various symmetrically inequivalent cation arrangements in perovskite oxides (Fig. \ref{fig:Main_ordering_dependence}). Specifically, when trained with unrelaxed perovskite structures and evaluated based on the $E_\mathrm{hull}$ of various cation arrangements relative to the ground state of the same composition, symmetry-invariant CGCNN shows an MAE of $18.8\ \mathrm{meV/atom}$ and an $R^2$ score of $-0.15$ between GCNN- and DFT-derived relative energies (Fig. \ref{fig:Main_ordering_dependence}b). Having a negative $R^2$ score for CGCNN indicates that this model performs more poorly than a baseline method assuming all symmetrically distinct cation arrangements (Fig. \ref{fig:Main_ordering_dependence}a) have the same energy. Switching the inputs of CGCNN from unrelaxed to relaxed structures does not improve results much (MAE: $17.5\ \mathrm{meV/atom}$; $R^2$ score: $0.11$; Fig. \ref{fig:Main_ordering_dependence}d). In contrast, e3nn performs more strongly when using unrelaxed (MAE: $14.1\ \mathrm{meV/atom}$; $R^2$ score: $0.37$; Fig. \ref{fig:Main_ordering_dependence}c) or relaxed structures as GCNN inputs (MAE: $12.0\ \mathrm{meV/atom}$; $R^2$ score: $0.55$; Fig. \ref{fig:Main_ordering_dependence}e). Overall, while invariant GCNNs fall short in capturing the relative thermodynamic stability of different atomic orderings in perovskite oxides, we found that equivariant GCNNs outperform invariant models in learning their thermodynamic stability against structural transformation into alternative orderings.

\begin{figure}
\phantomsection
\begin{center}
\includegraphics[max size={\textwidth}{\textheight}]{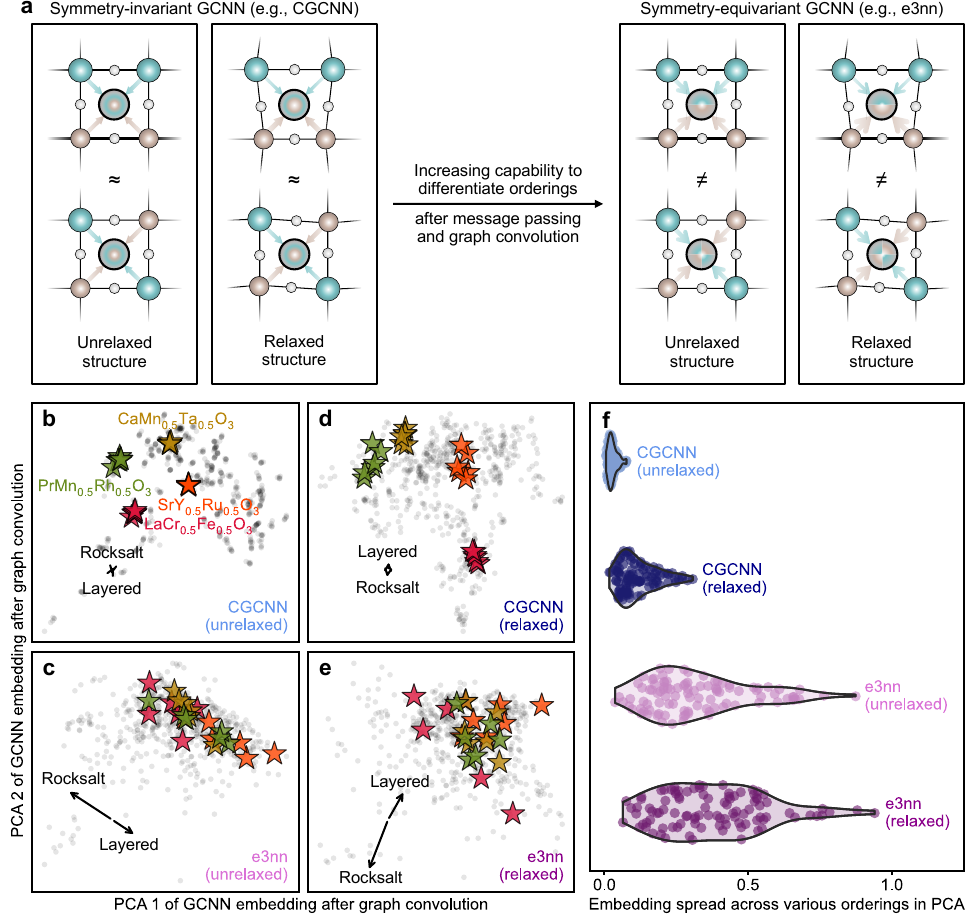}
\caption{
\textbf{Differentiating symmetrically inequivalent orderings by capturing crystallographic symmetries with GCNNs.}
\textbf{a}, Equivariant GCNNs can contrast the distinct symmetries of crystal structures with the same composition but different orderings more easily than their invariant counterparts. Color scheme: A, grey; B, cyan; B', brown; O, white. \textbf{b}--\textbf{f}, 2D PCA for the GCNN embeddings after graph convolution layers for all the six possible symmetrically distinct cation arrangements (Fig. \ref{fig:Main_ordering_dependence}a) of the holdout set of $100$ AB\textsubscript{0.5}B'\textsubscript{0.5}O\textsubscript{3} oxides. We trained CGCNN and e3nn on either unrelaxed (\textbf{b},\textbf{c}) or DFT-relaxed structures (\textbf{d},\textbf{e}) and further selected the best-performing model of each case for this latent embedding analysis, where the arrows show the embedding spread (multiplied by a factor of 2, for clarity) of an ordering relative to the mean of all orderings averaged over these $100$ perovskites. The embedding spread of different orderings for the same composition was quantified as the average Euclidean vector from each embedding to the centroid and further normalized by the spreads of all latent embeddings (\textbf{f}).
}
\label{fig:Main_ordering_rationale}
\end{center}
\end{figure}

The difference in performance across models can be interpreted through the quality and expressivity of the representations they learn (Fig. \ref{fig:Main_ordering_rationale}). We projected the latent embeddings of symmetrically distinct cation arrangements (Fig. \ref{fig:Main_ordering_rationale}a) onto a reduced two-dimensional (2D) representation via principal component analysis (PCA, Fig. \ref{fig:Main_ordering_rationale}b--e). In principle, more expressive models should embed these diverse orderings into distinct representations. As expected, DFT-relaxed structures create more differentiated embeddings, as they directly capture symmetry breaking in actual spatial coordinates, while message passing on unrelaxed structures cannot rely on distance-based edge features (e.g., bond lengths) and can only distinguish environment through the processing of node features (e.g., chemical elements) and orientation-based edge features (e.g., bond directions). Symmetry-invariant CGCNN learns very similar representations for the six symmetrically inequivalent cation orderings of AB\textsubscript{0.5}B'\textsubscript{0.5}O\textsubscript{3} (Fig. \ref{fig:Main_ordering_dependence}a) when trained on information-rich DFT-relaxed structures (Figs. \ref{fig:Main_ordering_rationale}d and \ref{fig:SI_embedding_analysis_cgcnn_relaxed}), and the invariant CGCNN captures essentially degenerate ordering representations for unrelaxed geometry inputs (Figs. \ref{fig:Main_ordering_rationale}b and \ref{fig:SI_embedding_analysis_cgcnn_unrelaxed}). On the contrary, for inequivalent atomic arrangements, symmetry-equivariant e3nn embeds the heterogeneous chemical environments into well-differentiated representations (Figs. \ref{fig:Main_ordering_rationale}c,e, \ref{fig:SI_embedding_analysis_e3nn_relaxed}, and \ref{fig:SI_embedding_analysis_e3nn_unrelaxed}), even when utilizing unrelaxed structures as inputs (Fig. \ref{fig:Main_ordering_rationale}c). Overall, the distribution of the latent embeddings for these six symmetrically unique cation orderings across various compositions (Fig. \ref{fig:Main_ordering_rationale}f) confirms that e3nn generally results in much more diverse latent embeddings for structures with identical compositions but distinct orderings, when compared with CGCNN. Moreover, for symmetry-equivariant models, one specific type of cation ordering (such as the rocksalt or layered ordering, Fig. \ref{fig:Main_ordering_rationale}b--e) generally shows up in a particular direction within the 2D PCA latent space relative to the other orderings of the same perovskite oxide composition. As learning different latent representations from inequivalent orderings is a prerequisite for GCNNs to capture the ordering dependence of materials properties, the behaviors of invariant and equivariant GCNNs (Fig. \ref{fig:Main_ordering_rationale}) can justify their performance in inferring the dependence of $E_\mathrm{hull}$ on various atomic orderings on materials properties (Fig. \ref{fig:Main_ordering_dependence}). Notably, in contrast to ordering dependence, altering the elemental identity of cations in these perovskites can readily give rise to drastically different latent embeddings via either symmetry-invariant or symmetry-equivariant message passing and graph convolution (Fig. \ref{fig:SI_compositional_rationale}). This sharp contrast renders the mastering of ordering-dependent materials properties a much more challenging task, when compared with learning their dependence on compositions. Future optimization of symmetry-aware, ordering-sensitive GCNNs needs systematic efforts that bridge the theory of deep representation learning for 3D geometric graphs with the fundamentals of complex symmetry relations in modern crystallography. Existing studies on graph representation learning for 3D atomic systems have focused on characterizing the expressive power of GCNNs by their ability to solve geometric graph isomorphism \cite{Pozdnyakov:2022,Joshi:2023}. The diversity of point and space groups in crystallography can result in various types of symmetry relations in different crystal systems \cite{Muller:2013}, which complicates the development of symmetry-aware GCNNs with ideal expressive power for diverse classes of crystallographic orderings.

\phantomsection
\addcontentsline{toc}{subsection}{Assessing the generalizability of ordering-sensitive graph neural networks}
\subsection*{Assessing the generalizability of ordering-sensitive graph neural networks}

\begin{figure}
\phantomsection
\begin{center}
\includegraphics[max size={\textwidth}{\textheight}]{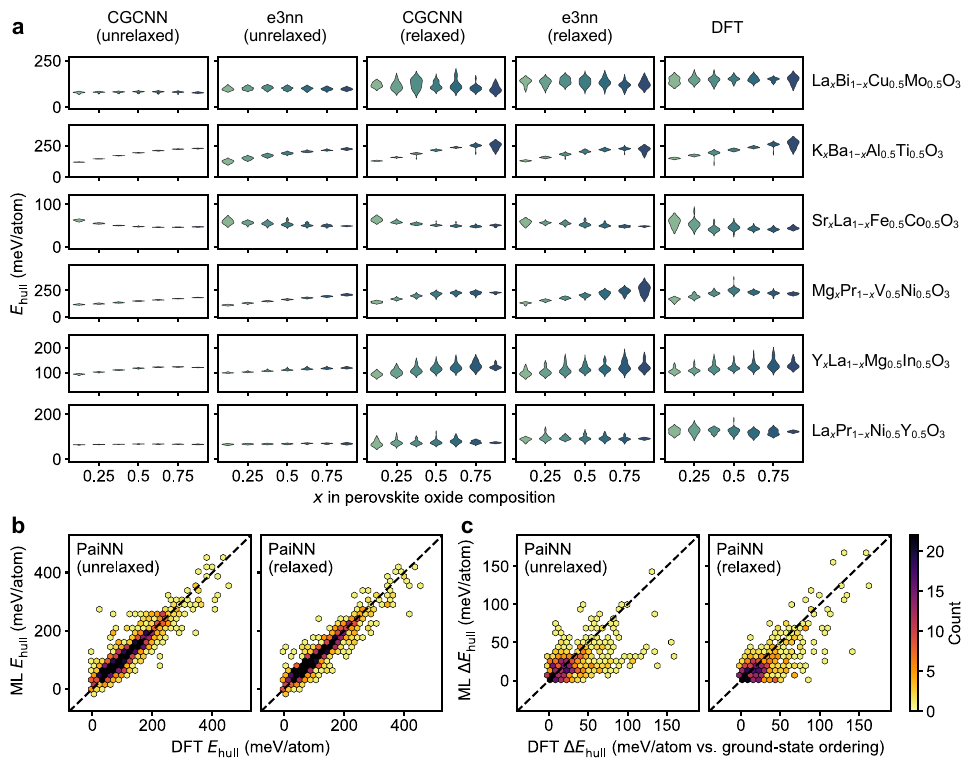}
\caption{
\textbf{Generalizability of ordering-sensitive GCNNs across model architectures and perovskite compositions.}
\textbf{a}, GCNN-predicted vs. DFT-computed $E_\mathrm{hull}$ for a holdout set of $863$ A\textsubscript{\textit{x}}A'\textsubscript{1--\textit{x}}B\textsubscript{0.5}B'\textsubscript{0.5}O\textsubscript{3} perovskite structures with various symmetrically inequivalent cation arrangements, where CGCNN and e3nn were trained on either unrelaxed or DFT-relaxed structures. Violin plots show the kernel density estimation of the energy distribution of these symmetrically distinct cation arrangements. \textbf{b}, PaiNN-predicted vs. DFT-computed $E_\mathrm{hull}$ for a test set of $1,261$ perovskites oxides. \textbf{c}, PaiNN-predicted vs. DFT-computed $E_\mathrm{hull}$ of all symmetrically distinct cation arrangements relative to those of PaiNN-predicted or DFT-computed ground-state arrangements, respectively, for a holdout set of $100$ AB\textsubscript{0.5}B'\textsubscript{0.5}O\textsubscript{3} perovskite oxide compositions, where each composition has six symmetrically inequivalent cation arrangements. PaiNN was trained on either unrelaxed or DFT-relaxed structures. All data points denote the mean of the three best-performing models.
}
\label{fig:Main_additional_generalizability}
\end{center}
\end{figure}

The capabilities of symmetry-equivariant GCNNs in differentiating atomic orderings are applicable to other materials classes and to a variety of model architectures. We focused on the ordering dependence of $E_\mathrm{hull}$ for AB\textsubscript{0.5}B'\textsubscript{0.5}O\textsubscript{3} (Figs. \ref{fig:Main_ordering_dependence} and \ref{fig:Main_ordering_rationale}) because of the abundant experimental literature \cite{King:2010} and the strong predictive power of DFT-computed $E_\mathrm{hull}$ to recapitulate experimental cation orderings \cite{Peng:2024}. For starters, we also evaluated model performance on the quinary A\textsubscript{\textit{x}}A'\textsubscript{1--\textit{x}}B\textsubscript{0.5}B'\textsubscript{0.5}O\textsubscript{3} perovskites (Fig. \ref{fig:Main_additional_generalizability}a), which can have correlated orderings between their A-site and B-site cations \cite{Peng:2024,King:2010}. Likewise, we chose e3nn as an archetypal state-of-the-art equivariant GCNN architecture with spherical tensors \cite{Batzner:2022,Batatia:2022,Merchant:2023,Batatia:2024,Chen:2021,Wen:2024,Okabe:2024}, but, for instance, another type of symmetry-equivariant GCNN architecture with Cartesian tensors, PaiNN \cite{Schutt:2021} (Fig. \ref{fig:Main_additional_generalizability}b,c), perform comparably to e3nn and much better than CGCNN (Figs. \ref{fig:Main_compositional_dependence} and \ref{fig:Main_ordering_dependence}). Future studies are needed to continue benchmarking the performance of symmetry-equivariant models in differentiating other categories of chemical (dis)order (e.g., defects) in various multicomponent materials (e.g., metal alloys) and further learning both the compositional and ordering dependence of diverse properties, such as catalytic activity \cite{Lunger:2024}, phonon dynamics \cite{Chen:2021,Okabe:2024}, and mechanical properties \cite{Wen:2024}.

\begin{figure}
\phantomsection
\begin{center}
\includegraphics[max size={\textwidth}{\textheight}]{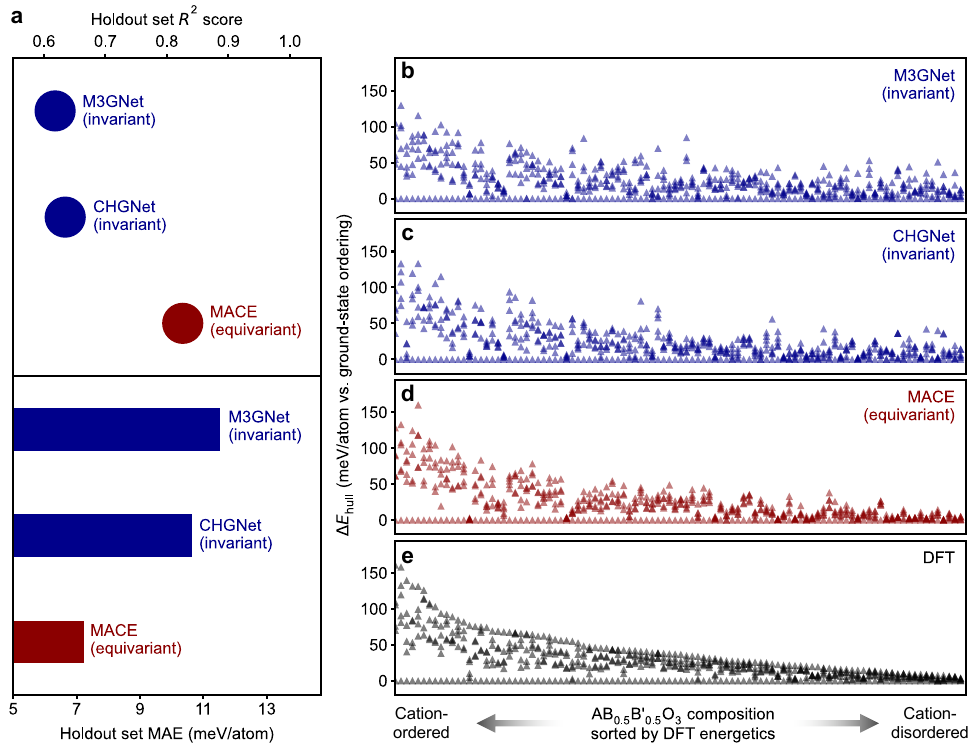}
\caption{
\textbf{Comparing symmetry-invariant and symmetry-equivariant GCNN interatomic potentials.}
\textbf{a}, $R^2$ score and MAE between the interatomic-potential-predicted vs. DFT-computed $E_\mathrm{hull}$ of all symmetrically unique cation arrangements relative to those of interatomic-potential-inferred or DFT-computed ground-state arrangements, respectively (Fig. \ref{fig:SI_dependence_interatomic_potentials}b), quantified for a holdout set of $100$ AB\textsubscript{0.5}B'\textsubscript{0.5}O\textsubscript{3} perovskite compositions, where each composition has six symmetrically distinct cation arrangements (Fig. \ref{fig:Main_ordering_dependence}a). \textbf{b}--\textbf{e}, M3GNet- (\textbf{b}), CHGNet- (\textbf{c}), and MACE-predicted (\textbf{d}) vs. DFT-computed $E_\mathrm{hull}$ (\textbf{e}) of the six symmetrically inequivalent cation arrangements with respect to ground-state arrangements, respectively, for each perovskite oxide in the holdout set. Oxide compositions are sorted based on the highest DFT-computed $E_\mathrm{hull}$ relative to the ground state in descending order. For these symmetrically distinct cation arrangements of AB\textsubscript{0.5}B'\textsubscript{0.5}O\textsubscript{3} perovskites (Fig. \ref{fig:Main_ordering_dependence}a), our previous work \cite{Peng:2024} has shown that having comparable DFT-computed energies gives rise to experimentally cation-disordered oxides, while having drastically different DFT-computed energies tends to result in experimentally cation-ordered ones.
}
\label{fig:Main_interatomic_potentials}
\end{center}
\end{figure}

The importance of symmetry equivariance in learning atomic orderings applies broadly to structure--property prediction tasks, including GCNN interatomic potentials that predict the total energies and, through automatic differentiation, inferring the atomic forces (Figs. \ref{fig:Main_interatomic_potentials} and \ref{fig:SI_dependence_interatomic_potentials}) needed for geometry relaxation and molecular dynamics simulations. We selected M3GNet \cite{Chen:2022} and CHGNet \cite{Deng:2023} as representative symmetry-invariant interatomic potentials that encode distance and angular information, whereas MACE \cite{Batatia:2022,Batatia:2024} was examined as a state-of-the-art symmetry-equivariant GCNN interatomic potential that captures the entire 3D crystallographic geometries. For multicomponent perovskite oxides, these three interatomic potentials have comparable performance in predicting composition-dependent thermodynamic stability (Fig. \ref{fig:SI_dependence_interatomic_potentials}a), and as they are used without modification, such interatomic potentials have prediction errors larger than custom-made GCNN prediction models based on CGCNN and e3nn (Fig. \ref{fig:Main_compositional_dependence}). In comparison, for inferring the energy dependence on atomic orderings, we found that GCNN interatomic potentials (Figs. \ref{fig:Main_interatomic_potentials}a and \ref{fig:SI_dependence_interatomic_potentials}b) generally perform better than direct prediction models (Fig. \ref{fig:Main_ordering_dependence}), presumably due to their capability to give rise to different trajectories of structural relaxation and energy minimization for two structures with inequivalent orderings even when their GCNN-predicted energies are similar prior to the structural optimization. More importantly, we observed that symmetry-equivariant MACE (Fig. \ref{fig:Main_interatomic_potentials}d) significantly outperforms its invariant counterparts (Fig. \ref{fig:Main_interatomic_potentials}b,c) for characterizing the ordering-dependent ab initio stability of perovskite oxides (Fig. \ref{fig:Main_interatomic_potentials}e), making MACE potentially capable of acting as a surrogate model to facilitate the a priori prediction of experimental orderings in these multicomponent materials based on the relative energetics of various symmetrically inequivalent cation arrangements \cite{Peng:2024}. Altogether, these results highlight that the strong expressive power and high prediction accuracy of symmetry-aware GCNNs for capturing orderings are not limited to direct prediction models and can be readily extended to universal interatomic potentials. Compared with direct prediction models, these interatomic potentials require orders of magnitude more expensive, ab initio data for model training \cite{Riebesell:2024} and behave orders of magnitude slower for model inference \cite{Okabe:2024}. Hence, apart from further engineering the architecture and strengthening the performance of symmetry-aware, ordering-sensitive GCNNs, future efforts should also be focused on coupling equivariant ML models with classical sampling methods or active learning approaches to facilitate efficient, rigorous navigation of high-dimensional atomic configuration spaces in complex materials.

\phantomsection
\addcontentsline{toc}{section}{Discussion}
\section*{Discussion}

In this study, we systematically examined an essential, yet often overlooked research question in the design of GCNNs for computational materials discovery---learning the dependence of key materials properties on both chemical compositions and atomic orderings simultaneously. We observed that although both symmetry-invariant and symmetry-equivariant GCNNs can characterize composition-dependent materials properties across the periodic table, equivariant GCNNs greatly outperform their invariant counterparts in predicting the dependence of materials properties on the different symmetrically inequivalent atomic orderings of the same composition. By visualizing how these models capture the chemical coordination environments in crystal structures through message passing and graph convolution, we demonstrated that only symmetry-equivariant models can effectively encode the crystallographic symmetries of crystal structures and utilize such information to identify the difference between inequivalent orderings. We found that the contrast between invariant and equivariant GCNNs in capturing orderings can be demonstrated for a wide variety of perovskite compositions and model architectures. Lastly, we pinpointed remaining challenges and promising paths forward for further optimizing the expressivity and generalizability of symmetry-equivariant, ordering-sensitive GCNNs.

Our findings underline the promise of symmetry-aware ML models in boosting efficient, yet rigorous materials and chemical discovery. A persistent concern from the experimental community about ML-driven computational materials discovery is that the various possibilities of chemical (dis)order are typically neglected \cite{Merchant:2023,Cheetham:2024}. Our recent work has demonstrated that the lack of consideration for diverse possible atomic orderings can result in significant sampling errors in assessing the formation energies and electronic-structure properties of multicomponent perovskite oxides with cation disorder through high-throughput computational screening \cite{Peng:2024}. To this end, symmetry-equivariant GCNNs can help the computational community take account of long- and short-range orders in multicomponent crystalline materials to ensure rigorous ML-enabled materials discovery. The capability to capture the energy dependence of these complex materials on their compositions and orderings concomitantly renders symmetry-aware GCNNs a powerhouse for efficiently gauging their thermodynamic stability against both phase decomposition into competing phases and structural transformation into alternative orderings. The explicit consideration of various chemical (dis)order can also make symmetry-equivariant GCNNs potentially more suitable than their invariant counterparts for driving the computational screening of multicomponent materials with desired properties that are exceptionally sensitive to atomic orderings, e.g., ferromagnetism and piezoelectricity. Lastly, we envision that the observed contrast between the symmetry-aware capability of equivariant and invariant GCNNs can be extended in future studies to the ML-powered chemical design of molecules and proteins, such as for uncovering the dependence of molecular reactivity on chirality and stereoisomerism by equivariant models.

\phantomsection
\addcontentsline{toc}{section}{Methods}
\section*{Methods}

\phantomsection
\addcontentsline{toc}{subsection}{High-throughput DFT calculations}
\subsection*{High-throughput DFT calculations}

A high-throughput computational dataset of structures and energies for multicomponent perovskite oxides was constructed through periodic plane-wave spin-polarized DFT calculations (Fig. \ref{fig:Main_knowledge_gap} and Table \ref{table:dataset_comparison}). We used an in-house automated DFT pipeline for structural optimization and energy calculation, using Perdew--Burke--Ernzerhof (PBE) functional \cite{Perdew:1996} as implemented in the Vienna Ab initio Simulation Package \cite{Kresse:1993,Kresse:1996}, projector augmented wave method \cite{Blochl:1994} for the description of core-electron interaction, and a plane-wave cutoff of $520\ \mathrm{eV}$. DFT calculations were conducted with Hubbard $U$ correction for V 3d ($U_{\mathrm{V}} = 3.25\ \mathrm{eV}$), Cr 3d ($U_{\mathrm{Cr}} = 3.7\ \mathrm{eV}$), Mn 3d ($U_{\mathrm{Mn}} = 3.9\ \mathrm{eV}$), Fe 3d ($U_{\mathrm{Fe}} = 5.3\ \mathrm{eV}$), Co 3d ($U_{\mathrm{Co}} = 3.32\ \mathrm{eV}$), Ni 3d ($U_{\mathrm{Ni}} = 6.2\ \mathrm{eV}$), Mo 4d ($U_{\mathrm{Mo}} = 4.38\ \mathrm{eV}$), and W 5d ($U_{\mathrm{W}} = 6.2\ \mathrm{eV}$) electrons, where the $U$ values were optimized by fitting the experimental formation enthalpies of binary oxides \cite{Wang:2006}. All these calculations were initialized with high-spin ferromagnetic states in order to assess a consistent and tractable set of magnetic structures. These DFT settings are compatible with the Materials Project database \cite{Jain:2013}, a good standard for high-throughput materials discovery. Notably, this compatibility with the Materials Project has also enabled us to computationally examine the thermodynamic stability (i.e., $E_\mathrm{hull}$) of multicomponent perovskite oxides by comparing the DFT energies of these perovskite oxides with competing phases in their DFT-computed compositional phase diagrams \cite{Bartel:2020}. While using more advanced density functional or evaluating more diversified initial magnetic configurations might improve the accuracy of DFT calculations, it is beyond the focus of this study and can be, in general, too computationally expensive for high-throughput studies. Furthermore, our recent work \cite{Peng:2024} has shown that, for hundreds of experimentally studied AB\textsubscript{0.5}B'\textsubscript{0.5}O\textsubscript{3} perovskites with various symmetrically unique cation arrangements, PBE-level DFT calculations with high-spin ferromagnetic initialization can be used to accurately infer the experimental cation orderings in these perovskite oxides, supporting the validity of DFT settings employed in this work.

For DFT structural optimization, at first, unrelaxed perovskite oxide structures were built in high-symmetry cubic supercells, with $40$ atoms and lattice parameters of $8\ \mathrm{\mathring{A}}$. The perovskite oxide compositions were selected from a list of $72$ elements (Fig. \ref{fig:Main_knowledge_gap} and Table \ref{table:dataset_comparison}), with stoichiometries of ABO\textsubscript{3}, A\textsubscript{\textit{x}}A'\textsubscript{1--\textit{x}}BO\textsubscript{3}, AB\textsubscript{\textit{y}}B'\textsubscript{1--\textit{y}}O\textsubscript{3}, and A\textsubscript{\textit{x}}A'\textsubscript{1--\textit{x}}B\textsubscript{\textit{y}}B'\textsubscript{1--\textit{y}}O\textsubscript{3}, where the ranges of $x$ and $y$ include $1$, $0.875$, $0.75$, $0.625$, $0.5$, $0.375$, $0.25$, $0.125$, and $0$. For a given oxide composition, the cation orderings in structures were sampled from a complete list of all possible symmetrically inequivalent cation arrangements in the cubic supercells. Our prior work \cite{Peng:2024} has shown that DFT structural relaxation from high-symmetry unrelaxed structures can hardly converge to ground-state atomic configurations with symmetry lower than idealized cubic \cite{Lufaso:2006} due to the too-high symmetry of DFT-computed forces for idealized cubic crystallographic environments. Therefore, in addition to this standard approach, we utilized a simple method \cite{Peng:2024} to break the high symmetry of unrelaxed perovskite structures. Specifically, prior to structural relaxation, we displaced all atoms in unrelaxed oxide structures by a small distance sampled from a uniform distribution with a range between $0.01\ \mathrm{\mathring{A}}$ and $ 0.1\ \mathrm{\mathring{A}}$. M3GNet \cite{Chen:2022} was also used as a pre-trained interatomic potential to pre-relax these structures before DFT structural relaxation. For DFT relaxation, the convergence threshold for electronic steps was $10^{-6}\ \mathrm{eV}$ per unit cell, and the residual forces on all atoms were lower than $10^{-2}\ \mathrm{eV\ \mathring{A}^{-1}}$. Similar to our recent work \cite{Peng:2024}, we found that breaking symmetry generally gives rise to DFT-relaxed structures with much lower energies than those without symmetry breaking, and thus, we took the lowest-energy structures to build our high-throughput DFT dataset.

\phantomsection
\addcontentsline{toc}{subsection}{GCNN training and inference}
\subsection*{GCNN training and inference}

We first selected CGCNN \cite{Xie:2018} and e3nn \cite{Geiger:2022} as representative ML models for symmetry-invariant and symmetry-equivariant GCNNs, respectively. We leveraged previous implementations from Xie et al. for CGCNN \cite{Xie:2018} and Chen et al. for e3nn \cite{Chen:2021}, respectively, to predict $E_\mathrm{hull}$ from unrelaxed or DFT-relaxed perovskite oxide structures. Notably, although the original e3nn implementation from Chen et al. uses a vector as the output to predict phonon density of states \cite{Chen:2021}, we modified this implementation by passing this vector through an additional series of fully connected layers to yield a scalar output as $E_\mathrm{hull}$. Moreover, the maximum order of spherical harmonics was set to $l = 2$. To fairly compare CGCNN and e3nn, we initialized the node features of both models using the same elemental initialization vectors from the original CGCNN implementation \cite{Xie:2018}.

For GCNN training and inference, we partitioned our perovskite oxide DFT dataset into training, validation, and test sets. As our work examines orderings in multicomponent perovskite oxides, all ABO\textsubscript{3} compositions were included in the training set. The training--validation--test split was conducted based on compositions, such that no specific oxide composition appeared in multiple sets, leading to $6,276$, $1,277$, and $1,261$ perovskite oxide structures in training, validation, and test sets, respectively. Furthermore, we built two holdout sets to evaluate whether CGCNN and e3nn can capture the ordering dependence of $E_\mathrm{hull}$ for perovskite oxide structures with the same composition but different cation orderings. The first holdout set has $600$ AB\textsubscript{0.5}B'\textsubscript{0.5}O\textsubscript{3} structures composed of $100$ different compositions with all six possible symmetrically inequivalent cation arrangements. Moreover, the second holdout set includes $863$ A\textsubscript{\textit{x}}A'\textsubscript{1--\textit{x}}B\textsubscript{0.5}B'\textsubscript{0.5}O\textsubscript{3} structures with various symmetrically distinct cation arrangements examined computationally. The perovskite compositions in the holdout sets are not present in the training, validation, or test sets. All training and inference were conducted with NVIDIA Volta V100 ($32$ GB), RTX 3080 ($10$ GB), RTX 2080 Ti ($11$ GB), and GTX 1080 Ti ($11$ GB) GPUs. The ranges of all the hyperparameters of CGCNN and e3nn are listed in Tables \ref{table:hyperparameter_cgcnn} and \ref{table:hyperparameter_e3nn}, respectively, and hyperparameter optimization was conducted using SigOpt \cite{Martinez-Cantin:2018} with a budget of 50 models. For a given set of hyperparameters, we trained the model for 100 epochs, where the best-performing iteration was saved according to performance on the validation set. After training, we extracted the GCNN embeddings of multicomponent perovskite oxides in CGCNN and e3nn by pooling node feature vectors and projecting these high-dimensional vectors linearly to a 2D space using PCA. The embedding spread of a single oxide composition in PCA was computed as the average Euclidean vector from each embedding of this composition to the centroid of all embeddings of such a composition. For a fair comparison between different GCNNs, we further normalized the embedding spreads of all oxide compositions under examination with respect to the average Euclidean vector from each of their GCNN embeddings to the centroid of all embeddings.

We additionally used PaiNN \cite{Schutt:2021} as a complementary example of symmetry-equivariant GCNNs. We modified a prior implementation of PaiNN for molecular property prediction from Axelrod et al. \cite{Axelrod:2022} by implementing periodic boundary conditions in message passing and graph convolution and initializing node features as the initialization vectors from the original CGCNN implementation \cite{Xie:2018}, with the ranges of hyperparameters listed in Table \ref{table:hyperparameter_painn}. Moreover, M3GNet \cite{Chen:2022}, CHGNet \cite{Deng:2023}, and MACE-MP-0 \cite{Batatia:2024} were leveraged as pre-trained interatomic potentials to test whether they have captured the energy dependence of multicomponent perovskite oxides as a function of compositions and orderings. Since these interatomic potentials aim to replicate the DFT-computed total energy of materials through structural optimization, $E_\mathrm{hull}$ can be further derived by constructing the energetic convex hull in compositional phase diagrams \cite{Bartel:2020}.

\phantomsection
\addcontentsline{toc}{section}{Data and code availability}
\section*{Data and code availability}

\noindent
Python computer code, DFT data, and trained models for reproducing this work are available via GitHub (https://github.com/learningmatter-mit/PerovskiteOrderingGCNNs), Zenodo (doi: 10.5281/zenodo.13820311) and Materials Data Facility (doi: 10.18126/ncqt-rh18). We used prior implementations of CGCNN \cite{Xie:2018}, e3nn \cite{Geiger:2022}, and PaiNN \cite{Axelrod:2022} with further modifications as submodules in our GitHub repository.

\phantomsection
\addcontentsline{toc}{section}{References}

\phantomsection
\addcontentsline{toc}{section}{Acknowledgments}
\section*{Acknowledgments}

\noindent
This project was funded by the Advanced Research Projects Agency–Energy, U.S. Department of Energy, under award number DE-AR0001220. This work used Expanse at San Diego Supercomputer Center through allocation MAT230007 from the Advanced Cyberinfrastructure Coordination Ecosystem: Services \& Support (ACCESS) program, which is supported by National Science Foundation grants 2138259, 2138286, 2138307, 2137603, and 2138296. This research used resources of the National Energy Research Scientific Computing Center (NERSC), a U.S. Department of Energy Office of Science User Facility located at Lawrence Berkeley National Laboratory, operated under Contract DE-AC02-05CH11231 using NERSC award m4074. The authors acknowledge the Engaging cluster at the Massachusetts Green High-Performance Computing Center and the SuperCloud cluster at the Lincoln Laboratory Supercomputing Center for providing high-performance computing resources.

\phantomsection
\addcontentsline{toc}{section}{Author contributions}
\section*{Author contributions}

\noindent
J.P., J.D., J.K., and R.G.-B. designed the studies. J.P., J.D., J.K., and J.R.L. performed the studies and analyzed the results. J.P. drafted the manuscript. All authors edited the manuscript. R.G.-B. supervised the project.

\phantomsection
\addcontentsline{toc}{section}{Competing interests}
\section*{Competing interests}

\noindent
The authors declare no competing interests.

\clearpage
\clearpage
\baselineskip18pt

\setcounter{page}{1}

\stepcounter{myequation}
\renewcommand{\theequation}{S\arabic{equation}}

\begin{center}

\vspace{36pt}{\Large Supplementary Information for}\\[12pt]

{\bf\large Learning Ordering in Crystalline Materials with Symmetry-Aware Graph Neural Networks}

\vspace{6pt} \hspace{0mm}Jiayu Peng,$^{1,\dag}$ James Damewood,$^{1,2,\dag}$ Jessica Karaguesian,$^{1,2,\dag}$ Jaclyn R. Lunger,$^{1}$ Rafael Gómez-Bombarelli$^{1,*}$\\

\vspace{6pt} $^{\dag}$\small Equal contribution; $^{*}$\small rafagb@mit.edu (R.G.-B.)

\end{center}

\pagebreak

\clearpage
\newcounter{sfigure}
\renewcommand{\figurename}{Fig.}
\renewcommand{\thefigure}{S\arabic{sfigure}}

\addtocounter{sfigure}{1}
\begin{figure}
\phantomsection
\begin{center}
\includegraphics[max size={\textwidth}{\textheight}]{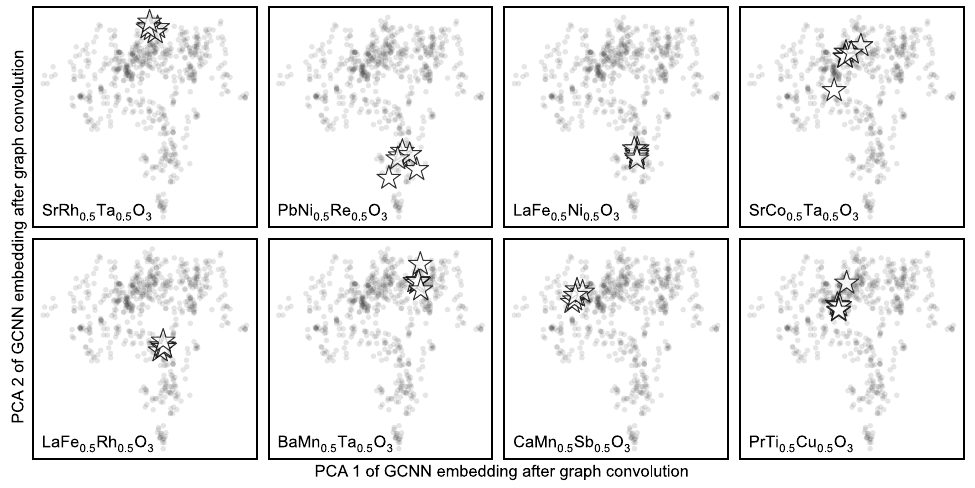}
\caption{
\textbf{Embeddings of symmetrically inequivalent DFT-relaxed structures in CGCNN.}
}
\label{fig:SI_embedding_analysis_cgcnn_relaxed}
\end{center}
\end{figure}

\addtocounter{sfigure}{1}
\begin{figure}
\phantomsection
\begin{center}
\includegraphics[max size={\textwidth}{\textheight}]{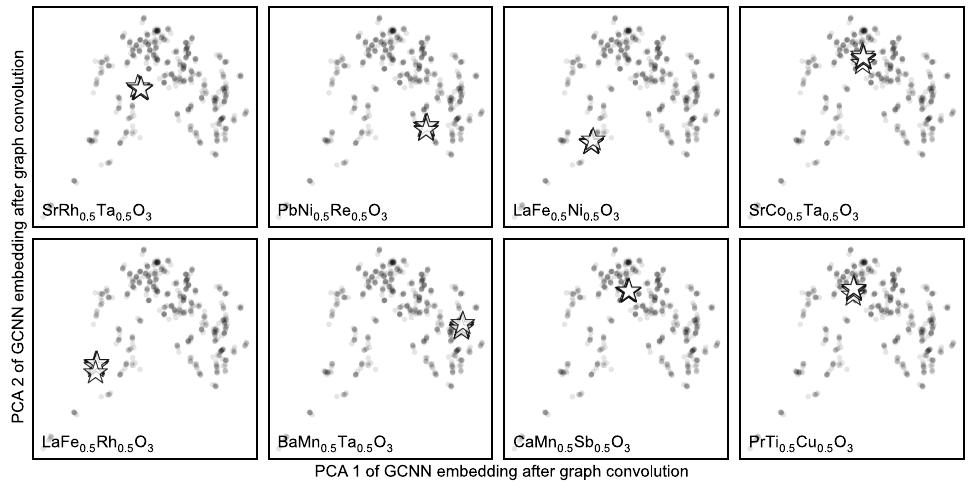}
\caption{
\textbf{Embeddings of symmetrically inequivalent unrelaxed structures in CGCNN.}
}
\label{fig:SI_embedding_analysis_cgcnn_unrelaxed}
\end{center}
\end{figure}

\addtocounter{sfigure}{1}
\begin{figure}
\phantomsection
\begin{center}
\includegraphics[max size={\textwidth}{\textheight}]{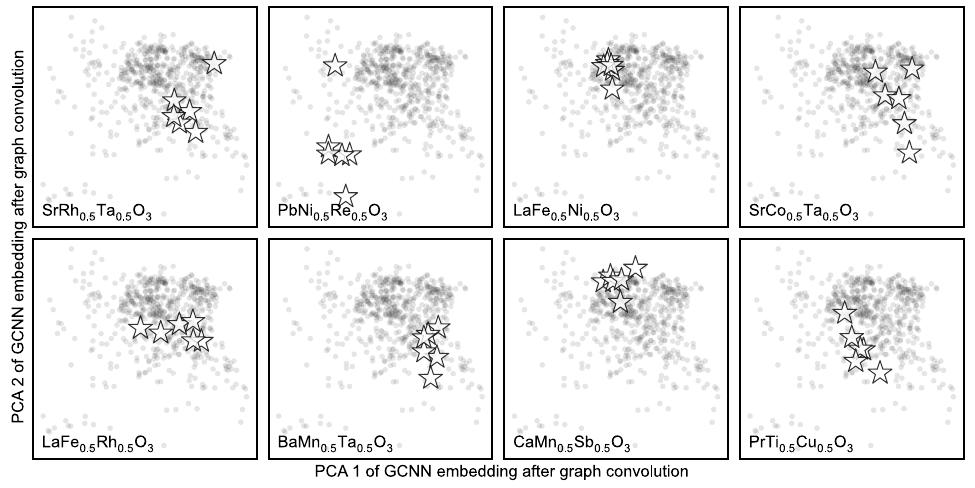}
\caption{
\textbf{Embeddings of symmetrically inequivalent DFT-relaxed structures in e3nn.}
}
\label{fig:SI_embedding_analysis_e3nn_relaxed}
\end{center}
\end{figure}

\addtocounter{sfigure}{1}
\begin{figure}
\phantomsection
\begin{center}
\includegraphics[max size={\textwidth}{\textheight}]{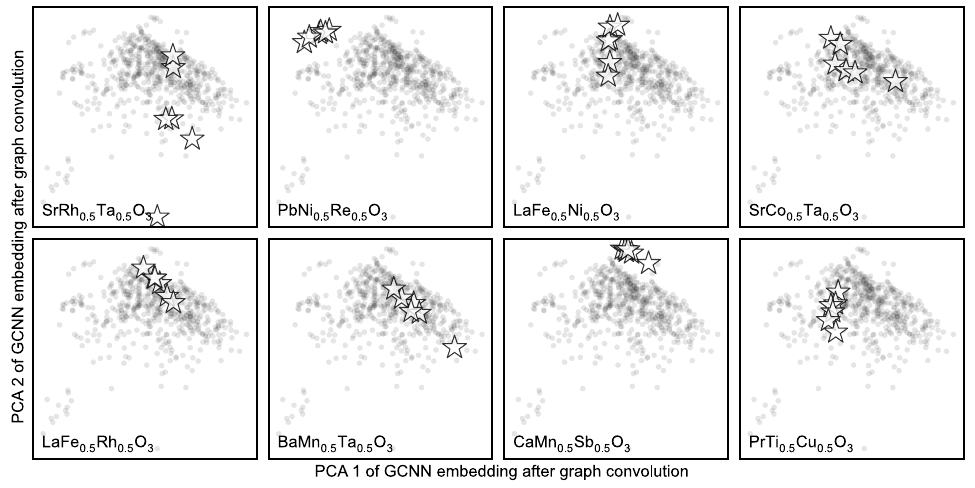}
\caption{
\textbf{Embeddings of symmetrically inequivalent unrelaxed structures in e3nn.}
}
\label{fig:SI_embedding_analysis_e3nn_unrelaxed}
\end{center}
\end{figure}

\addtocounter{sfigure}{1}
\begin{figure}
\phantomsection
\begin{center}
\includegraphics[max size={\textwidth}{\textheight}]{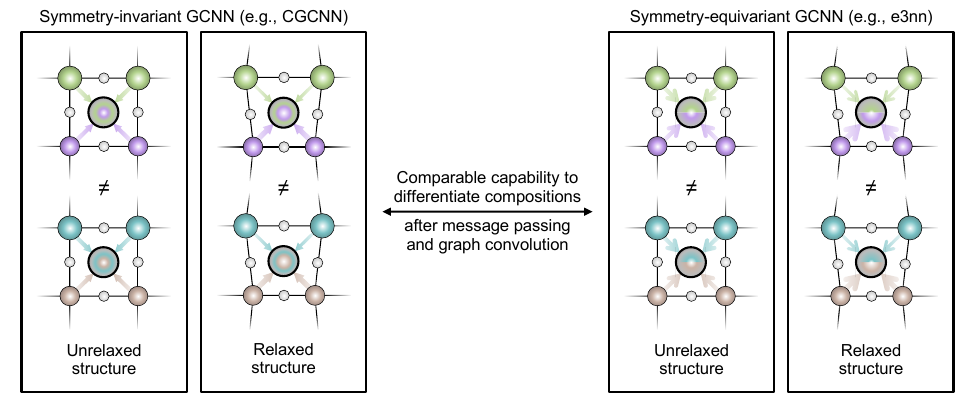}
\caption{
\textbf{Capturing the compositional dependence of key materials properties via message passing and graph convolution is much easier than learning the ordering dependence (Fig. \ref{fig:Main_ordering_rationale}a) for both equivariant and invariant GCNNs.}
}
\label{fig:SI_compositional_rationale}
\end{center}
\end{figure}

\addtocounter{sfigure}{1}
\begin{figure}
\phantomsection
\begin{center}
\includegraphics[max size={0.95\textwidth}{\textheight}]{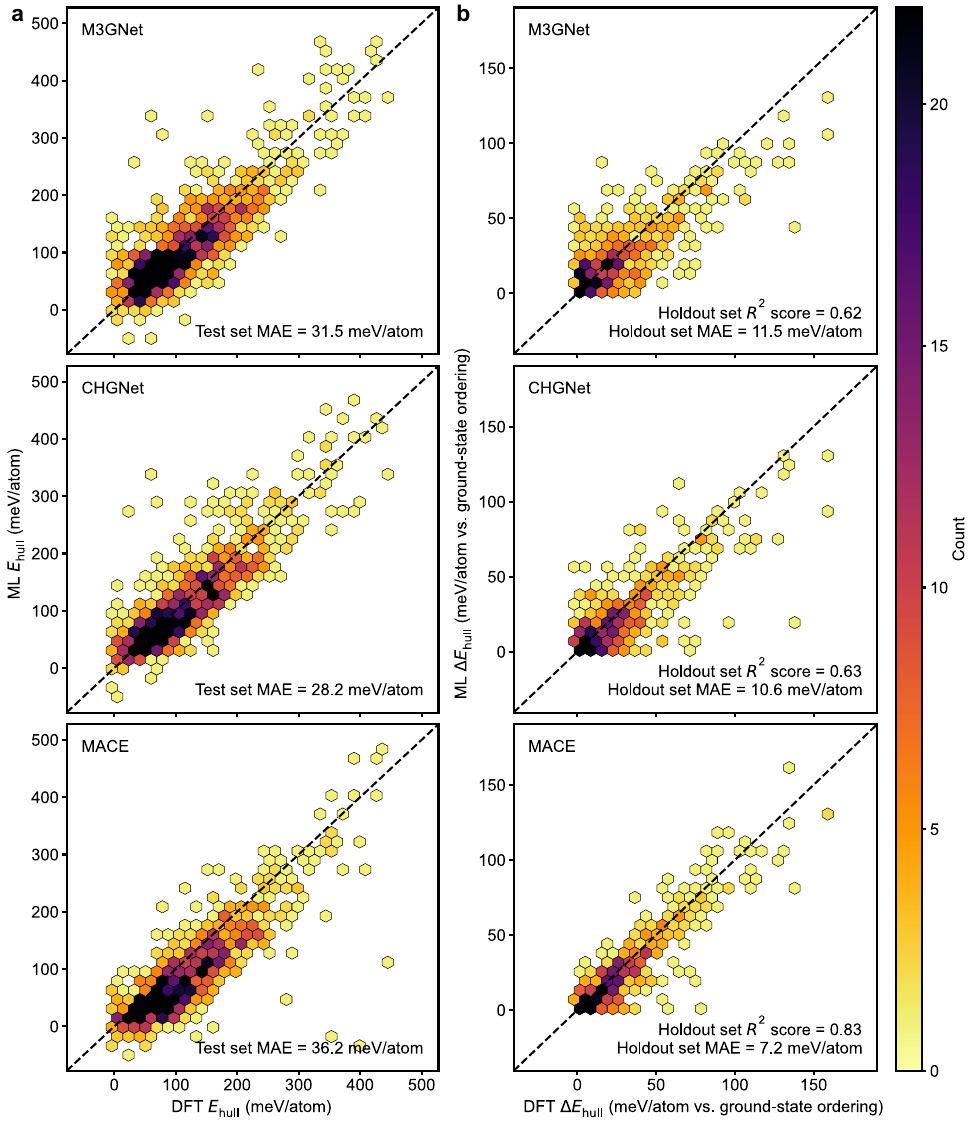}
\caption{
\textbf{State-of-the-art universal interatomic potentials.}
\textbf{a}, Interatomic-potential-predicted vs. DFT-computed $E_\mathrm{hull}$ for a test set of $1,261$ perovskites. \textbf{b}, Interatomic-potential-predicted vs. DFT-computed $E_\mathrm{hull}$ of all symmetrically inequivalent cation arrangements relative to those of interatomic-potential-inferred or DFT-computed ground-state arrangements, respectively, for a holdout set of $100$ AB\textsubscript{0.5}B'\textsubscript{0.5}O\textsubscript{3} compositions, where each composition has six symmetrically inequivalent cation arrangements.
}
\label{fig:SI_dependence_interatomic_potentials}
\end{center}
\end{figure}

\clearpage
\newcounter{SItable}
\renewcommand{\tablename}{Table}
\renewcommand{\thetable}{S\arabic{SItable}}

\addtocounter{SItable}{1}
\begin{ThreePartTable}
\centering
\begin{xltabular}{\linewidth}{llrrlr}

\caption{
\textbf{Comparison between the DFT dataset in this work and recent high-throughput studies that have examined at least thousands of perovskite oxides.}
}
\label{table:dataset_comparison}\\

\toprule
Dataset & Ordering & Element & Total & Stoichiometry & Count \\
\midrule
\endfirsthead

\caption[]{(continued from previous page)} \\
\toprule
Dataset & Ordering & Element & Total & Stoichiometry & Count \\
\midrule
\endhead

\bottomrule
\endfoot

\bottomrule
\insertTableNotes
\endlastfoot

& & & & ABO\textsubscript{3} & $2,659$ \\
This work & Various & $72$ & $10,277$ & A\textsubscript{\textit{x}}A'\textsubscript{1--\textit{x}}BO\textsubscript{3}, AB\textsubscript{\textit{y}}B'\textsubscript{1--\textit{y}}O\textsubscript{3} & $2,328$ \\
& & & & A\textsubscript{\textit{x}}A'\textsubscript{1--\textit{x}}B\textsubscript{\textit{y}}B'\textsubscript{1--\textit{y}}O\textsubscript{3} & $5,290$ \\
\hline
Ref. \cite{Castelli:2012} & Single & $52$ & $2,704$ & ABO\textsubscript{3} & $2,704$ \\
\hline
Ref. \cite{Emery:2017} & Single & $73$ & $5,329$ & ABO\textsubscript{3} & $5,329$ \\
\hline
Ref. \cite{Korbel:2016} & Single & $55$ & $>32,000$ & ABX\textsubscript{3} (mostly non-oxides) & $>32,000$ \\
\hline
Ref. \cite{Schmidt:2017} & Single & $64$ & $249,654$ & ABX\textsubscript{3} (mostly non-oxides) & $249,654$ \\
\hline
& & & & ABO\textsubscript{3} & $72$ \\
Ref. \cite{Jacobs:2018} & Single & $42$ & $2,145$ & A\textsubscript{\textit{x}}A'\textsubscript{1--\textit{x}}BO\textsubscript{3}, AB\textsubscript{\textit{y}}B'\textsubscript{1--\textit{y}}O\textsubscript{3} & $1,359$ \\
& & & & A\textsubscript{\textit{x}}A'\textsubscript{1--\textit{x}}B\textsubscript{\textit{y}}B'\textsubscript{1--\textit{y}}O\textsubscript{3} & $714$ \\
\hline
Ref. \cite{Talapatra:2021} & Single & $68$ & $3,469$ & ABO\textsubscript{3}, A\textsubscript{0.5}A'\textsubscript{0.5}B\textsubscript{0.5}B'\textsubscript{0.5}O\textsubscript{3} & $3,469$ \\
\hline
Ref. \cite{Ma:2021} & Single & $\geq 42$ & $2,913$ & ABO\textsubscript{3}, AB\textsubscript{0.5}B'\textsubscript{0.5}O\textsubscript{3} & $2,913$ \\
\hline
Ref. \cite{Wang:2022} & Single & $14$ & $2,401$ & Sr\textsubscript{\textit{x}}A\textsubscript{1--\textit{x}}Fe\textsubscript{\textit{y}}B\textsubscript{1--\textit{y}}O\textsubscript{3--$\delta$} & $2,401$ \\
\hline
& & & & A\textsubscript{0.5}A'\textsubscript{0.5}BO\textsubscript{3} & $2,980$ \\
Ref. \cite{Bare:2023} & Single & $39$ & $66,516$ & AB\textsubscript{0.5}B'\textsubscript{0.5}O\textsubscript{3} & $3,679$ \\
& & & & A\textsubscript{0.5}A'\textsubscript{0.5}B\textsubscript{0.5}B'\textsubscript{0.5}O\textsubscript{3} & $59,857$ \\
\hline
Ref. \cite{Wang2:2024} & Single & $54$ & $4,900$ & AB\textsubscript{0.5}B'\textsubscript{0.5}O\textsubscript{3} & $4,900$ \\

\end{xltabular}

\end{ThreePartTable}
\pagebreak

\addtocounter{SItable}{1}
\begin{ThreePartTable}
\centering
\begin{xltabular}{\linewidth}{lr}

\caption{
\textbf{Hyperparameter space of CGCNN.}
}
\label{table:hyperparameter_cgcnn}\\

\toprule
Hyperparameter & Range \\
\midrule
\endfirsthead

\caption[]{(continued from previous page)} \\
\toprule
Hyperparameter & Range \\
\midrule
\endhead

\bottomrule
\endfoot

\bottomrule
\insertTableNotes
\endlastfoot

Length of node features & $32$--$256$ \\
Number of convolutional layers & $2$--$5$ \\
Length of hidden features & $32$--$256$ \\
Number of hidden layers & $1$--$4$ \\
\hline
Batch size & $4$--$16$ \\
Adam optimizer learning rate & $10^{-5}$, $10^{-4}$, $10^{-3}$, $10^{-2}$ \\
\begin{tabular}{@{}l@{}} Number of epochs with no improvement \\ after which learning rate will be reduced \end{tabular} & $10$--$30$ \\

\end{xltabular}

\end{ThreePartTable}
\pagebreak

\addtocounter{SItable}{1}
\begin{ThreePartTable}
\centering
\begin{xltabular}{\linewidth}{lr}

\caption{
\textbf{Hyperparameter space of e3nn.}
}
\label{table:hyperparameter_e3nn}\\

\toprule
Hyperparameter & Range \\
\midrule
\endfirsthead

\caption[]{(continued from previous page)} \\
\toprule
Hyperparameter & Range \\
\midrule
\endhead

\bottomrule
\endfoot

\bottomrule
\insertTableNotes
\endlastfoot

Length of node features & $32$, $64$, $128$ \\
Number of convolutional layers & $1$--$4$ \\
Multiplicity of irreducible representation & $16$, $32$, $64$ \\
Number of basis for radial filters & $5$, $10$, $20$ \\
\begin{tabular}{@{}l@{}} Number of neurons in the hidden layers \\ of the radial fully connected network \end{tabular} & $32$, $64$, $128$ \\
Length of hidden features & $32$, $64$, $128$ \\
Number of hidden layers & $0$--$2$ \\
\hline
Batch size & $4$--$12$ \\
Adam optimizer learning rate & $10^{-5}$, $10^{-4}$, $10^{-3}$, $10^{-2}$ \\
\begin{tabular}{@{}l@{}} Number of epochs with no improvement \\ after which learning rate will be reduced \end{tabular} & $10$--$30$ \\

\end{xltabular}

\end{ThreePartTable}
\pagebreak

\addtocounter{SItable}{1}
\begin{ThreePartTable}
\centering
\begin{xltabular}{\linewidth}{lr}

\caption{
\textbf{Hyperparameter space of PaiNN.}
}
\label{table:hyperparameter_painn}\\

\toprule
Hyperparameter & Range \\
\midrule
\endfirsthead

\caption[]{(continued from previous page)} \\
\toprule
Hyperparameter & Range \\
\midrule
\endhead

\bottomrule
\endfoot

\bottomrule
\insertTableNotes
\endlastfoot

Length of node features & $32$, $64$, $128$, $256$, $512$ \\
Number of convolutional layers & $1$--$6$ \\
Activation function & \begin{tabular}{@{}r@{}} ReLU, leaky ReLU, \\ swish, parametric swish \end{tabular} \\
\hline
Batch size & $4$--$16$ \\
Adam optimizer learning rate & $10^{-5}$, $10^{-4}$, $10^{-3}$ \\
\begin{tabular}{@{}l@{}} Number of epochs with no improvement \\ after which learning rate will be reduced \end{tabular} & $10$--$30$ \\

\end{xltabular}

\end{ThreePartTable}
\pagebreak

\end{document}